\documentclass[prd,nofootinbib,reprint,superscriptaddress]{revtex4-2}

\usepackage[utf8]{inputenc}
\usepackage{graphicx,color} 
\usepackage{xcolor}
\usepackage{rotating}
\usepackage{amssymb}
\usepackage{amsfonts}
\usepackage{amsmath}
\usepackage{xspace}
\usepackage{mathtools} 
\usepackage{verbatim}
\usepackage{comment}
\usepackage{enumitem}
\usepackage{physics}
\usepackage[colorlinks]{hyperref}
\usepackage{MnSymbol}
\usepackage{lineno}

\begin{document}
\title{Discriminating between different modified dispersion relations from gamma-ray observations} 

\author{S.~Caroff}
\email{sami.caroff@lapp.in2p3.fr}
\affiliation{Laboratoire d'Annecy de physique des particules, Université Savoie Mont-Blanc, CNRS/IN2P3, \\9 Chem. de Bellevue, 74940 Annecy}

\author{C.~Pfeifer}
\email{christian.pfeifer@zarm.uni-bremen.de}
\affiliation{ZARM, University of Bremen, Am Fallturm 2
, 28359 Bremen, Germany.}

\author{J.~Bolmont}
\affiliation{Sorbonne Universit\'e, CNRS/IN2P3,\\ Laboratoire de Physique Nucl\'eaire et de Hautes Energies, LPNHE,\\ 4 Place Jussieu, F-75005 Paris, France}

\author{T.~Terzi\'c}
\affiliation{University of Rijeka, Faculty of Physics,\\ Radmile Matejčić 2, 51000 Rijeka, Croatia}

\author{A.~Campoy-Ordaz}
\affiliation{Departament de F\'isica, Universitat Aut\`onoma de Barcelona and CERES-IEEC,\\ E-08193 Bellaterra, Spain}

\author{D.~Kerszberg}
\affiliation{Sorbonne Universit\'e, CNRS/IN2P3,\\ Laboratoire de Physique Nucl\'eaire et de Hautes Energies, LPNHE,\\ 4 Place Jussieu, F-75005 Paris, France}

\author{M.~Martinez}
\affiliation{Institut de F\'isica d'Altes Energies (IFAE),\\ The Barcelona Institute of Science and Technology (BIST), E-08193 Bellaterra (Barcelona), Spain}

\author{U.~Pensec}
\affiliation{Sorbonne Universit\'e, CNRS/IN2P3,\\ Laboratoire de Physique Nucl\'eaire et de Hautes Energies, LPNHE,\\ 4 Place Jussieu, F-75005 Paris, France}

\author{C.~Plard}
\affiliation{Laboratoire d'Annecy de physique des particules, Université Savoie Mont-Blanc, CNRS/IN2P3, \\9 Chem. de Bellevue, 74940 Annecy}

\author{J.~Stri\v{s}kovi\'c}
\affiliation{Josip Juraj Strossmayer University of Osijek, Department of Physics,\\ 31000 Osijek, Croatia}

\author{S.~Wong}
\affiliation{Physics Department, McGill University, Montreal,\\ QC H3A 2T8, Canada}

\begin{abstract}
The fact that the standard dispersion relation for photons in vacuum could be modified because of their interaction with the quantum nature of spacetime has been proposed more than two decades ago. A quantitative model [Jacob \& Piran, JCAP 01, 031 (2008)], has been tested extensively using distant highly energetic astrophysical sources, searching for energy-dependent time delays in photon arrival times. Since no delay was firmly measured, lower limits were set on the energy scale $\Lambda$ related to these effects.
In recent years, however, different but equally well-grounded expressions beyond the Jacob \& Piran model were obtained for the photon dispersion relation, leading to different expressions for the dependence of lag versus redshift. 
This article introduces a general parameterization of modified dispersion relations in homogeneous and isotropic, i.e. cosmological, symmetry which directly leads to a general parameterized lag versus redshift dependence encompassing both existing and new models. This parameterization could be used in the future to compare the predicted time lags of the different models and test them against observations. 
To investigate this possibility, realistic data sets are simulated, mimicking different types of extragalactic sources as detected by current and future instruments. When no lag is injected in the simulated data, each lag-redshift model leads, as expected, to a different value for the limit on $\Lambda$, and the Jacob \& Piran model gives the most stringent bound. When a lag at $\Lambda \sim E_P$ in the Jacob \& Piran model is injected, it is detected for all the other lag-redshift relations considered, although leading to different values. Finally, the possibility to discriminate between several lag-redshift models is investigated, emphasizing the importance of an evenly distributed sample of sources across a wide range of redshifts.
\end{abstract}

\maketitle

\section{Introduction} \label{sec:intro}
A most prominent feature in the effective description of the propagation of photons on a quantum spacetime is an energy-dependent time of arrival, which may be interpreted as an energy-dependent velocity. Photons of different energies emitted simultaneously at the same point in spacetime would be detected on Earth with a time delay, which emerges from the interaction of photons with quantum properties of gravity (see the recent reviews \cite[\S~5.1]{Addazi:2021xuf}, \cite[\S~3.8]{Amelino-Camelia:2008aez} and references therein for a detailed discussion about the emergence of the effect). The structures which cause such an effect could be gravitons~\cite{Kempf:1994su,Das:2010zf,Garattini:2011kp}, non-commutative geometry~\cite{AmelinoCamelia:2011cv}, string theory backgrounds~\cite{Ellis:2008gg}, spin-foam~\cite{Gambini:1998it}, or one of the many other approaches considered in the construction of quantum spacetime from a fundamental theory of quantum gravity (QG). Since a self-consistent fundamental theory of QG is still elusive and since the derivation of the interaction of photons with QG is a difficult task, effective methods are used. Modified dispersion relations (MDRs) are often employed to model the propagation of photons on quantum spacetime. This can be understood analogously to the description of photons propagating through a medium. Fundamentally, photons interact with all constituents of the medium described by the local Lorentz-invariant standard model of physics. Effectively, this propagation can be described by a medium-dependent non-Lorentz invariant dispersion relation which captures the interaction of photons with the medium, leading to dispersive and refractive effects. Such an effective description of photon propagation on quantum spacetime is an important approach to quantum gravity phenomenology \cite{Addazi:2021xuf,AlvesBatista:2023wqm}. The appearing deviations from local Lorentz invariance, whether they come from a Lorentz invariance violation (LIV) model or from deformations of special relativity (DSR), may or may not be present in the fundamental theory, as the analogy with the propagation of photons in a medium nicely demonstrates. 

In the context of searches for QG in astrophysics, time delays are derived from different models by solving the point particle equations of motion for photons subject to a homogeneous and isotropic, i.e.\ cosmologically symmetric, dispersion relation \cite{Ellis:2002in,Jacob:2008bw,Rosati:2015pga,Barcaroli:2016yrl,Pfeifer:2018pty,Amelino-Camelia:2020bvx,Amelino-Camelia:2023srg}. The absence of any delay, or a significant detection thereof, immediately leads to constraints on the effective models, and gives a guideline for the semi-classical effective limits of fundamental theories of quantum gravity. It has to be pointed out though, that in case a significant delay would be measured, it would still be necessary to ascertain whether its origin is due to QG effects or to another competing phenomena, for example source-intrinsic delays, or mechanisms leading to new physics beyond general relativity and the standard model of particle physics.

Despite the expectation that a QG-induced time delay will be very small, long travel distances and high photon energies act as amplifiers of the effect, and may bring it into our observational capabilities. The search for time delays focuses on gamma-ray observations of sources with as large a redshift as possible. 
Energy-dependent time delays have been extensively investigated by various gamma-ray experiments at high ($E>100$\,MeV) and very-high energies ($E>100$\,GeV), in particular with space-borne experiments like \textit{Fermi}\,\footnote{\url{https://fermi.gsfc.nasa.gov}}~\cite{Fermi-LAT:2009ihh, 2009ApJ...702..791M} or the Neil Gehrels \textit{Swift} observatory\,\footnote{\url{https://swift.gsfc.nasa.gov}}~\cite{SwiftScience:2004ykd}, ground-based imaging atmospheric Cherenkov telescopes (IACTs), such as H.E.S.S.\,\footnote{\url{https://www.mpi-hd.mpg.de/hfm/HESS/}}~\cite{2015arXiv150902902H}, MAGIC\,\footnote{\url{https://magic.mpp.mpg.de}}~\cite{MAGIC:2014zas}, VERITAS\,\footnote{\url{https://veritas.sao.arizona.edu}}~\cite{Park:2015ysa}, and the first CTAO\,\footnote{\url{https://www.ctao.org/}}~\cite{CTAObservatory:2022mvt} telescope \mbox{LST-1}\,\footnote{\url{https://www.lst1.iac.es/}}~\cite{CTA-LSTProject:2023haa}, as well as ground-based wide-field hybrid observatories such as LHAASO\,\footnote{\url{http://english.ihep.cas.cn/lhaaso/}}~\cite{Ma:2022aau}.

Until now, no lag was measured at these energies, and limits have been set on the relevant quantum gravity or new physics energy scale, called $\Lambda$ in the following, which suppresses the effect. It is assumed that, at that scale, the influence of quantum spacetime, or presently unknown physical phenomena beyond general relativity and the standard model become dominant, and the perturbative treatment breaks down. While limits on $\Lambda$ can exceed the Planck scale $E_P \sim 10^{19}$ GeV for individual gamma-ray bursts (GRBs), e.g. \cite{Vasileiou_2013,LHAASO:2024lub}, it is only of the order of $10^{17}$~GeV when several \textit{Fermi} GRBs are combined \cite{Ellis2019}. The strongest constraints obtained with flaring active galactic nuclei (AGN) are of the order of $2\times 10^{18}$ GeV, e.g. \cite{Abramowski2011, Abe_2024, Terzi__2021}.

Since there are many MDR models proposed in the literature, each predicting different time delays, the goal of this article is to probe the different predictions with realistic simulated data sets and to compare the resulting bounds on the QG scale. The time delay formula that is derived from the most general perturbation of the general relativistic dispersion relation of photons on Friedmann-Lema\^itre-Robertson-Walker (FLRW) spacetime had been obtained in \cite{Pfeifer:2018pty}. As it turns out, currently, it is not possible to analyze this general expression against simulated or real data. Therefore, we particularize the general time delay to parameterized MDRs that are polynomial in the momenta. By choosing the parameters, this model covers numerous, if not all, MDRs discussed in the literature as well as new ones to leading order beyond general relativity. The advantage of this parameterized approach is that it sets up a framework with which MDRs can be tested against time delay observations. In case of the detection of a significant time delay in the future, the appearing parameters will serve as fitting parameters to find the most suitable MDR compatible with the observation.

To find the bounds on the QG scale from different MDRs, we recall how to understand homogeneous and isotropic MDRs as Hamilton functions and how to derive the time delay in Sec.~\ref{ssec:general}. We then present a new systematic parameterization of MDRs and their resulting time delay in Sec.~\ref{ssec:parammod}.
For the analysis of simulated datasets, seven MDR models are selected, two of them leading to an identical time delay at first order. Five of them have already been discussed in the literature already (Jacob \& Piran \cite{Jacob:2008bw}, $\kappa$-Poincaré in the bicrossproduct basis~\cite{Barcaroli:2015xda}, curvature-induced DSR~\cite{Amelino-Camelia:2020bvx}, and two FLRW-DSR \cite{Amelino-Camelia:2023srg} models), while the other two models are new alternatives to the Jacob \& Piran or the $\kappa$-Poincaré on FLRW spacetime model, which we discuss here for the first time. In Sec.~\ref{sec:dataanalysis1}, the simulations and the data analysis procedure are described. The results are then given and discussed in Sec.~\ref{sec:results}. We conclude the article with the summary of our results and their outlook in Sec.~\ref{sec:DiscOut}.

\section{Time delays from modified dispersion relations} \label{sec:theory}

In order to derive the time delay between photons of different energies emitted in the same spacetime event, we interpret dispersion relation as Hamilton functions on the single particle phase space of spacetime (technically the cotangent bundle) and derive their trajectories on a homogeneous and isotropic universe. In this section, we recall the mathematical details for general MDRs beyond general relativity (\ref{ssec:general}), we present an extended, very general, parameterization of MDRs, which cover all models that are studied in the literature (\ref{ssec:parammod}). Then, we discuss the explicit examples (\ref{ssec:MDRsspecific}) which will be compared to simulated data in Sec.~\ref{sec:dataanalysis1}.

We would like to highlight that the parameterization of MDRs is chosen such that they can be easily implemented in an analysis code. Thus, given any MDR or time delay formula, one simply has to specify the parameter functions for the model under consideration and the comparison with observational or simulated data can be performed directly.

\subsection{The general time delay formula}\label{ssec:general}

Dispersion relations on curved spacetimes are encoded in Hamilton functions on the single particle phase space of spacetime \cite{Barcaroli:2015xda}. The Hamilton function determines the dispersion relation by equating it to a constant, usually called $m^2$ (where $m$ is interpreted as the inertial mass of the particle), and determines the motion of the point particle via the Hamilton equations of motion ($a=0,1,2,3$):

\begin{align}
    \mathcal{H}(x,p) = m^2\,,\quad \dot p_a = -\partial_{x^a} \mathcal{H}\,,\quad \dot x^a = \partial_{p_a} \mathcal{H}\,. 
\end{align}

To study the propagation of gamma rays from cosmological distances to telescopes on Earth or in Earth orbit, we assume a homogeneous and isotropic dispersion relation, which reflects the large-scale symmetry of our universe. The most general dispersion relation of this kind is encoded in the Hamiltonian \cite{Pfeifer:2018pty}

\begin{eqnarray}
	\mathcal{H}(x,p) & = & \mathcal{H}(t,p_t, w),\nonumber\\ 
 w^2& =& p_r^2 \chi^2 + \frac{p_\theta^2}{r^2} + \frac{p_\phi^2}{r^2 \sin^2\theta}\,,
\end{eqnarray} 
where $(r, \theta, \phi)$ are the usual spherical spatial coordinates employed in homogeneous and isotropic symmetry, $\chi = \sqrt{1-k r^2}$, $k$ being the spatial curvature of spacetime, and $w$ the total spatial momentum of the particle. Due to the high symmetry, the Hamilton equations of motion, which determine the propagation of the particle through spacetime, can partly be solved explicitly and reduce to

\begin{minipage}{0.15\textwidth} 
	\begin{align}
	\dot p_t&=- \partial_t \mathcal{H}\,,\label{eq:dotpt}  \\
    p_r&=\frac{w}{\chi} \label{eq:pr} \,, \\
	p_\theta &= 0\,,\\
	p_\phi&=0\,,
	\end{align}
\end{minipage}
\hspace{30pt}
\begin{minipage}{0.15\textwidth}
	\begin{align}
	\dot t &= \partial_{p_t} \mathcal{H}\, \label{eq:dott}, \\
	\dot r & = \chi \partial_wH   \, \label{eq:dotr},\\
	\theta & = \frac{\pi}{2}\,,\\
	\phi &= 0\,.
	\end{align}
\end{minipage}\vspace{11pt}

In order to determine the time delay induced by MDRs emerging from quantum gravity, we consider perturbations of the general relativistic dispersion relation. These are described by Hamilton functions of the type

\begin{align}\label{eq:Hpert}
	\mathcal{H}(t,p_t,w) =  p_t{}^2 - a(t)^{-2} w^2 +  h(t,p_t,w)\,,
\end{align}
where in the following all results are derived to first order in $h$. For $h = 0$, the expression leads to the Hamilton function that determines the motion of test particles in FLRW spacetime geometry.

The derivation of the general time delay formula for this kind of dispersion relation for massless particles has been presented in~\cite{Pfeifer:2018pty}. First, the massless dispersion relation $\mathcal{H}(x,p) = 0$ is solved for the energy of the particles $p_t=E(t,w)$ (measured by observers at rest in the chosen cosmological coordinate system). Then, one determines the redshift $z+1 = p_t(t_0)/p_t(t_1)$ and solves the Hamilton equation of motion for the radial motion to determine $r(t,E(t,w))$. Let us consider two photons, labeled $A$ and $B$, emitted at the same time from the same source with energies $E_A$ and $E_B$, respectively. The observed time delay can be deduced from the times $t_A$ and $t_B$ taken by the photons to reach the observer, that is a point in spacetime such that $r(t_A,E_A(t_A,w)) = r(t_B,E_B(t_B,w))=R$.
After some calculations, which are presented in detail in~\cite{Pfeifer:2018pty}, the time delay formula is obtained to first order in $\epsilon$ as

\begin{align}\label{eq:DT}
    \Delta t(z) = \int_{0}^{z}dz'\ \frac{f(z',w_2) - f(z',w_1)}{H(z')}\,,
\end{align}
where $H(z)$ is the Hubble parameter, hereafter taken as $H(z) = H_0 \sqrt{\Omega_m\,(z+1)^3 + \Omega_\Lambda}$, neglecting radiation density and curvature density. The function $f$ takes the form 

\begin{eqnarray}
    f(z,w) &=& \frac{1}{2 (p_t^0)^2} \bigg[ h(t,p_t^0,w) - p_t^0 \partial_{p_t} h(t,p_t^0,w) \nonumber\\
    & & -\ w \partial_w h(t,p_t^0,w)  \bigg]\,,
\end{eqnarray}
with $p_t^0 = w/a(t)$ being the zeroth order general relativistic dispersion relation for massless particles. The $t$-dependence of $h$ is assumed to be such that it can be replaced by the general relativistic (zeroth order) redshift

\begin{align}
    \label{eq:cosmRedshift}
    z(t) + 1 = \frac{a(t_0)}{a(t)} = \frac{1}{a(t)}\,,
\end{align}
meaning that we assume that this equation can be inverted for a dependence $t = t(z)$. We set the value of the scale factor today $a(t_0)$ to $1$. To first order in a quantum gravity or new physics energy scale $\Lambda$ one might expand the function $f$ which defines the general time delay formula \eqref{eq:DT} as $f(z,w) = (a w/\Lambda)\,f(z)$ to obtain 
\begin{align}\label{eq:DTlambda}
    \Delta t(z) = \frac{p_{t2}^0-p_{t1}^0}{\Lambda}\int_{0}^{z}dz'\ \frac{f(z')}{H(z')}\,.
\end{align}
However, such a generic formula, which includes a free function $f(z)$, is currently not suitable to be tested against data.

In order to be able to develop a systematic scheme to scrutinize the leading order deviations from general relativity of MDRs, we will now present a parameterization for perturbations of the general relativistic dispersion relation \eqref{eq:Hpert}, which is suitable to represent many models discussed in the literature. Most importantly, this parameterized approach is designed in a way that it is practical to be implemented into analysis codes. For this type of dispersion relations, we give an explicit expression for the time delay equation \eqref{eq:DT}. 

\subsection{Parameterizing modified dispersion relations}\label{ssec:parammod}

In order to understand the emergence of the time delay from the MDR on curved spacetime we start by specifying the modification of the MDR defining Hamilton function. On curved spacetime, this is an observer-independent and coordinate-invariant way to implement MDRs. As discussed previously, a homogeneous and isotropic modified dispersion must be a function of $p_t$ and $w$. Hence, the most general MDR of polynomial type in these variables, normalized by the scale $\Lambda$ at which the effects of the MDR become relevant is
\begin{align}\label{eq:MDRpertGen1}
    h(z,p_t,w) = p_t^2 \sum_q \sum_m \frac{1}{\Lambda^q}\ A_{qm}(t(z))\ p_t^{m} w^{q-m}\,,
\end{align}
where the summation indices $q, m$ are integers which can take positive and negative values, depending on the model of interest. In general, the coefficients $A_{qm}(t(z))$ can be functions of the time coordinate $t$, or in particular of the scale factor $a(t)$, and its derivatives. We always assume that it is possible to change the dependence from time $t$ to redshift $z$. In the context of MDRs emerging from quantum gravity, $\Lambda$ may be of the order of the Planck energy $E_P$. It is often denoted by $E_{QG}$ in most articles describing experimental searches for MDRs with astrophysical sources. The goal is to measure or constrain this parameter explicitly from data collected by gamma-ray observatories.

Choosing the parameter functions $A_{qm}$ allows to consider different powers of $\Lambda$ as lowest order correction to the general relativistic dispersion relation. For example, for the $\kappa$-Poincaré dispersion relation in the bicrossproduct basis on curved spacetime \cite{Barcaroli:2016yrl}, to second order one has 
\begin{align}\label{eq:kappa2}
       h(t,p_t,w) 
       = \frac{1}{\Lambda} \frac{p_t w^2}{a^2}
       + \frac{1}{\Lambda^2} \frac{(a^2 p_t^4 - 6 p_t^2 w^2)}{12 a^2}\,, 
\end{align}
and thus, at this order, the non-vanishing coefficients are $A_{1,-1} = 1/a^2$, $A_{2,2} = 1/12$ and $A_{2,0} =-1/(2a^2) $.
For the search of an imprint of the dispersion relation on gamma-ray observations, we fix the lowest order $I$ of $\Lambda$ to be the relevant correction to the general relativistic dispersion relation\,\footnote{Note that in many experimental papers the order $I$ is denoted by the letter $n$.}. This is done by choosing $A_{qm} = \delta_{qI} B_m$, where $\delta_{qI} = 0$ for $q\neq I$ and $\delta_{qI}=1$ for $I=q$, which leads to
\begin{align}\label{eq:MDRpertGen2}
    h(z,p_t,w) = p_t^2 \frac{1}{\Lambda^I} \sum_m \ B_{m}(t(z))\ p_t^{m} w^{I-m}\,.
\end{align}
Different models are now selected by specifying the parameter function $B_m$. Numerous MDRs discussed in the literature, as well as new ones, can be obtained and later be analyzed by choosing the $B_m$ for example as follows:
\begin{itemize}
    \item Setting $B_{m} = \delta_{Im}$ gives the famous Jacob \& Piran model \cite{Jacob:2008bw},
    \begin{align}
        h(t,p_t,w) = p_t^2 \frac{p_t^{I}}{\Lambda^I}\,.
    \end{align}
    \item Setting $I=1$ and $B_{m}=\delta_{m-1}/a^2$ gives the first, leading, order $\kappa$-Poincaré dispersion relation on FLRW spacetime in the bicrossproduct basis, see \eqref{eq:kappa2},
    \begin{align}\label{eq:kappa2}
       h(t,p_t,w) 
       = \frac{1}{\Lambda} \frac{p_t w^2}{a^2}\,. 
    \end{align}
    \item Setting $I=1$ and $B_{m}=\delta_{m-1}$ gives a new model that is rescaled with respect to the $\kappa$-Poincaré on FLRW spacetime model. We include this example in order to demonstrate how different MDRs lead to different redshift dependencies of the time delay. Moreover, the comparison of this model with the $\kappa$-Poincaré model shows how the different redshift dependencies influence the time delay prediction, which in turn influences the resulting bounds on $\Lambda$ from the data analysis
    \begin{align}
       h(t,p_t,w) = \frac{1}{\Lambda} p_t w^2\,. 
    \end{align}
    \item Setting $I=1$ and  $B_{m}= (D_s \delta_{ms} + C_v \delta_{mv-1})$ for fixed $s$ and $v$ yields a general mixed powers in $p$ and $w$ model, which have been studied very little so far,
    \begin{align}
        h(t,p_t,w) = p_t^2 \frac{1}{\Lambda} ( D_s p_t^s w^{1-s} + C_v p_t^{1-v} w^v)\,.
    \end{align}
    Among all of these possible mixed models, specifying $s=1,v=0$ and $D_{1} = - C_0 = \lambda$ (or equivalently $s=-1, v=2$ and $D_{-1} = - C_2 = -\lambda$), leads to the dispersion relation that was considered in \cite[Eq. (3)]{Amelino-Camelia:2020bvx} in $E(p)$ form, see Eq.~\eqref{eq:E(p)1}. Since we want to include such dispersion relations in the data analysis we already present this model here.
    \item Setting $B_{m}= \delta_{m0}$ gives a Jacob \& Piran type model, where the dispersion relation is modified by a power series of the spatial momenta instead of the energy
    \begin{align}\label{eq:SpaM0}
        h(t,p_t,w) = p_t^2 \frac{w^I}{\Lambda^I}\,.
    \end{align}
    To our knowledge, this model has not yet been discussed in the literature. We consider it interesting since, as we will see in the next section, it leads to the same time delay that originally emerged in \cite{Ellis:2005sjy} at the first order $I=1$.
\end{itemize}
Hence, for newly emerging MDRs in the future, it is possible to analyze them with the framework we present in the following of this article, by just determining the model parameters $B_m$. The other way around, many more models can be constructed by different choices of $B_m$. They may be explored in separate papers.

To continue, we solve the massless dispersion relation $\mathcal{H}(x,p)=0$ for the energy of the photon for an observer at rest $p_t=E$ as function of the spatial momentum $w=c p$ and convert the dispersion relation \eqref{eq:MDRpertGen2} into its more common form 
\begin{align}\label{eq:E(p)1}
    E 
    =\frac{p c}{a}\left(1 
    - \frac{1}{2} \frac{1}{\Lambda^I} \sum_m \left[\ B_m \ \frac{(p c)^I}{a^m} \right] \right)\,. 
\end{align}
We would like to highlight that the most general polynomial modification of the general relativistic dispersion relation at a given order $\Lambda^I$ \eqref{eq:MDRpertGen2}, leads to a sum over all possible orders of the scale factor $a(t)$. The $B_m$, in general, may be arbitrary functions of time. Using the expression for the cosmological redshift Eq.~\eqref{eq:cosmRedshift} and adopting a certain cosmological model, this dependence on time can be translated to dependence on the redshift $z$. 
In principle, $B_m$ could depend on the Hubble parameter, its derivatives, the spacetime curvature, or any other physical field on spacetime. These possible dependencies are a feature that emerges already at first $\Lambda$ order $I=1$. In general, this will lead to a more abundant phenomenology as for the specific models that have been studied in the literature. With this parametric \textit{ansatz} we systematically study the different possibilities to modify the photon dispersion relations, beyond the original suggestions \cite{Ellis:2002in,RodriguezMartinez:2006ee, Jacob:2008bw}, to be able to incorporate all possible polynomial models, of which there are many, as already became clear from the ones listed above.

Next, we will see explicitly what kind of expression for the time delays the different terms imply.

\subsection{The time delay formula for polynomial type perturbative modified dispersion relations}
\label{ssec:MDRsspecific}

We can now evaluate the time delay formula \eqref{eq:DT} for the general $I$-th order model \eqref{eq:MDRpertGen2} and obtain
\begin{align}\label{eq:DT2}
    \Delta t = & \frac{(I+1)}{2}\frac{(E_{01}^I-E_{02}^I)}{\Lambda^I}\nonumber\\  & \times \int_0^z \sum_m \frac{B_m(t(z'))(z'+1)^m}{H(z')}dz'\,,
\end{align}
where we choose $E_{01}>E_{02}$. Here, we see the full power of the parameterized approach. For a fixed $\Lambda$-order $I$, there is still freedom for the dependence of the MDR on the redshift through the choice of the parameter functions $B_m$, which are determined by the modification of the general relativistic dispersion relation from which we started. Moreover, this choice also determines whether the time delay between high-energy and low-energy photons being emitted at a certain redshift is positive or negative. We refer to these behaviors as subluminal (high-energy photon arrives after the low-energetic one) or superluminal (high-energy photon arrives before the low energy one), respectively. 

From Eq.~\eqref{eq:DT2} we can construct a generic geometric light distance function 
\begin{align}\label{eq:kappaB}
    \kappa(z) = \int_0^z \sum_m \frac{B_m(t(z'))(z'+1)^m}{H(z')}dz'\,.
\end{align}
The long term goal is to reconstruct $ \kappa(z)$, i.e. the values of $B_m(t(z))$, from observations such that it fits the data and the distribution of available sources over a wide redshift range. However, this would require a statistically significant detection of a time lag.

For the different specific models mentioned earlier we find from Eq.~\eqref{eq:DT2} the following predictions by specifying $I$ and inserting the non-vanishing coefficients $B_m$. The boxed equations below are those that will be used for the analysis in Sec.~\ref{sec:dataanalysis1}. The labels for these equations will be used later on to refer to each particular model.
\begin{itemize}
    \item $B_m = \delta_{Im}$ gives the famous Jacob \& Piran result~\cite{Jacob:2008bw},
    \begin{align}
       \Delta t = \frac{(I+1)}{2}\frac{(E_{01}^I-E_{02}^I)}{\Lambda^I} \int_0^z \frac{(z'+1)^I}{H(z')}dz' \,. 
    \end{align}
    For data analysis, we consider only the cases $I=1$: 
    \begin{align}\label{eq:JP}
        \boxed{
           \Delta t = \frac{(E_{01}-E_{02})}{\Lambda} \int_0^z \frac{(z'+1)}{H(z')}dz' \,, \tag{JP} }
    \end{align} 
    while the order $I=2$ gives
    \begin{align}\label{eq:JP2}
       \Delta t = \frac{3}{2}\frac{(E_{01}^2-E_{02}^2)}{\Lambda^2} \int_0^z \frac{(z'+1)^2}{H(z')}dz' \,.
    \end{align}
    \item $I=1$ and $B_{m}=\delta_{m-1}/a^2 = \delta_{m-1}(z+1)^2$ gives the result for the first order $\kappa$-Poincaré dispersion relation on FLRW spacetime in the bicrossproduct basis \cite{Barcaroli:2016yrl}. At first order, the resulting time delay is identical to the one of the Jacob \& Piran model \eqref{eq:JP}. The reason is that at first order the  $E(p)$ form \eqref{eq:E(p)1} of the  Jacob \& Piran and $\kappa$-Poincaré model take the same form. This degeneracy of the $E(p)$ form of these two dispersion relations is not present at higher orders.
    \item $I=1$ and $B_{m}=\delta_{m-1}$ gives the result for the rescaled $\kappa$-Poincaré dispersion relation on FLRW spacetime. Since there are no further parameters to fix, we can directly use it for the data analysis
    \begin{align}\label{eq:RekaP}
       \boxed{\Delta t = \frac{(E_{01}-E_{02})}{\Lambda} \int_0^z \frac{(z'+1)^{-1}}{H(z')}dz' \,. \tag{RekaP}}
    \end{align}
    \item  $I=1$ and  $B_{m}= (D_s \delta_{ms} + C_v \delta_{mv-1})$ for fixed $s$ and $v$ implies
    \begin{align}
        \Delta t = & \frac{(E_{01}-E_{02})}{\Lambda}\nonumber\\ 
         & \times\int_0^z \frac{D_s(z'+1)^s + C_v (z'+1)^{v-1}}{H(z')}dz' \,.
    \end{align}
    Choosing $s=1,v=0$ and $D_{1} = - C_0 = \lambda = 1$ (or equivalently $s=-1, v=2$ and $D_{-1} = - C_2 = -\lambda = 1$) gives the curvature induced time delay formula that was derived in \cite{Amelino-Camelia:2020bvx}, and which we will use later
    \begin{align}\label{eq:curvInd}
        \boxed{\Delta t 
        =  \frac{(E_{01}-E_{02})}{\Lambda}\ \int_0^z \frac{2z'+z'^2}{H(z')(z'+1)}dz' \,. \tag{CInd}}
    \end{align}
    \item $B_m = \delta_{m0}$ gives a Jacob \& Piran type model, where the dispersion relation is modified by a power series of the spatial momenta, instead of the energy. This implies
    \begin{align}
         \Delta t = \frac{(I+1)}{2}\frac{(E_{01}^I-E_{02}^I)}{\Lambda^I} \int_0^z \frac{1}{H(z')}dz' \,.
    \end{align}
    As for the original Jacob \& Piran model, we will only consider the case $I=1$
    \begin{align}\label{eq:SpaM}
        \boxed{
           \Delta t = \frac{(E_{01}-E_{02})}{\Lambda} \int_0^z \frac{1}{H(z')}dz' \,. \tag{SpaM} }
    \end{align} 
    The case $I=2$ gives the following expression for the delay:
    \begin{align}\label{eq:NewJP2}
       \Delta t = \frac{3}{2}\frac{(E_{01}^2-E_{02}^2)}{\Lambda^2} \int_0^z \frac{1}{H(z')}dz' \,.
    \end{align}
    The time delay \eqref{eq:SpaM} is similar to the one of the~\eqref{eq:JP} formula, except for the term $(1+z)$ in the integral which is not present here, since the derivation started from a different modification of the dispersion relation\,\footnote{Interestingly, the expression \eqref{eq:SpaM} is the same as the one originally obtained in \cite{Ellis:2005sjy}. In the latter article, published in 2007, it turned out that the factor $(1+z)$ was overlooked in the derivation of the formula \eqref{eq:JP} which would be introduced by Jacob \& Piran a year later \cite{Jacob:2008bw}.}.
    \item $I=1$ and non-vanishing $B_m$
    \begin{align}
        B_1 &= \eta_1\,,\\
        B_0 &= 2(\eta_2 + 2 \eta_3) H(z') I(z')\,, \\
        B_{-1} &= -(\eta_2+6\eta_3) H(z')^2 I(z')^2\,,\\
        B_{-2} &= 4 \eta_3 H(z')^3 I(z')^3\,,\\
        B_{-3} &= -\eta_3 H(z')^4 I(z')^4\,,
    \end{align}
    gives the models of \cite{Amelino-Camelia:2023srg}, parameterized by $\eta_1, \eta_2$ and $\eta_3$:
    \begin{widetext}
    \begin{equation}
        \Delta t  = \frac{(E_{01}-E_{02})}{\Lambda}
        \int dz'\Biggl(\frac{(z'+1)}{H(z')} 
         \left[\eta_1+\eta_2 \left(1-\left(1-\frac{H(z') I(z')}{z'+1}\right)^2\right)+\eta_3 \left(1-\left(1-\frac{H(z') I(z')}{z'+1}\right)^4\right)\right]\Biggr)\,,
    \end{equation}
    \end{widetext}
    where the function $I(z')$ is given by
    \begin{align}
        I(z') = \int_0^{z'} \frac{dz''}{H(z'')}\,. 
    \end{align}
    The special cases of \cite{Amelino-Camelia:2023srg} are obtained by the cor\-responding choices of parameters $\eta_1, \eta_2$ and $\eta_3$: 
    \begin{itemize}
        \item The choice $\eta_1=\eta_2=0,\ \eta_3=1$ leads to \cite[Eq. (4.7)/Fig. 1]{Amelino-Camelia:2023srg}, Eq.~\eqref{eq:DSR1} below;
        \item When $\eta_1=0,\ \eta_2=4,$ and $\eta_3=-3$, we obtain the following relation \cite[Eq. (4.8)/Fig. 2]{Amelino-Camelia:2023srg}, Eq.~\eqref{eq:DSR2}:
    \end{itemize}
\end{itemize}

        \begin{widetext}
        \begin{align}\label{eq:DSR1}
            \boxed{\Delta t  = \frac{(E_{01}-E_{02})}{\Lambda}
            \int dz'\Biggl(\frac{(z'+1)}{H(z')} 
            \left(1-\left(1-\frac{H(z') I(z')}{z'+1}\right)^4\right)\Biggr)\,; \tag{DSR1}}
        \end{align}
        \begin{align}\label{eq:DSR2}
        \boxed{\Delta t  = \tfrac{(E_{01}-E_{02})}{\Lambda}
        \int dz'\left(\tfrac{(z'+1)}{H(z')} 
         \left[4 \left(1-\left(1-\tfrac{H(z') I(z')}{z'+1}\right)^2\right)-3 \left(1-\left(1-\tfrac{H(z') I(z')}{z'+1}\right)^4\right)\right]\right)\,. \tag{DSR2}}
        \end{align}
        \end{widetext}

To summarize, the time delay formulas we use for the analysis in the following are \eqref{eq:JP}, \eqref{eq:SpaM}, \eqref{eq:RekaP}, \eqref{eq:curvInd}, \eqref{eq:DSR1} and \eqref{eq:DSR2}. We discovered that to leading first order, there exist different MDRs, which lead to the same expression for the time delay. Hence, there is a degeneracy between models, and their time delay predictions. This, and the infinite number of models one can construct and analyze in principle, make it necessary to have a clear guidance from theoretical investigations how to interpret and how to select models for the data analysis. The other way around, as soon as there is a measurement of a relevant time delay, it can be used to fit the coefficients $B_m$, and identify the optimal choice, which in turn selects a viable MDR.

The equations \eqref{eq:JP}, \eqref{eq:SpaM}, \eqref{eq:RekaP}, \eqref{eq:curvInd}, \eqref{eq:DSR1} and \eqref{eq:DSR2} are expressed in a way that the only free parameter left is the quantum gravity scale $\Lambda$. Also, using Eq.~\eqref{eq:kappaB}, all these equations can be written as
\begin{equation}
\label{eq:DtKappa}
\Delta t = \frac{(E_{01}-E_{02})}{H_0\ \Lambda} \times \kappa_*(z),
\end{equation}
with $I = 1$ and where $\kappa_*(z) = \kappa(z)\ H_0$. Here the Hubble constant $H_0$ is factored out so that $\kappa_*$ is dimensionless. The values for cosmological parameters are taken from~\cite[Table 2.1]{PhysRevD.110.030001}. For the Hubble constant, we use the value $H_0 = 73\,\mathrm{km\,s}^{-1}\,\mathrm{Mpc}^{-1}$ from the distance ladder measurements, because the sources in our studies are located at $z<1$. 

\begin{figure}[!b]
	\centering
   	  \includegraphics[width=3.4in, angle=90]{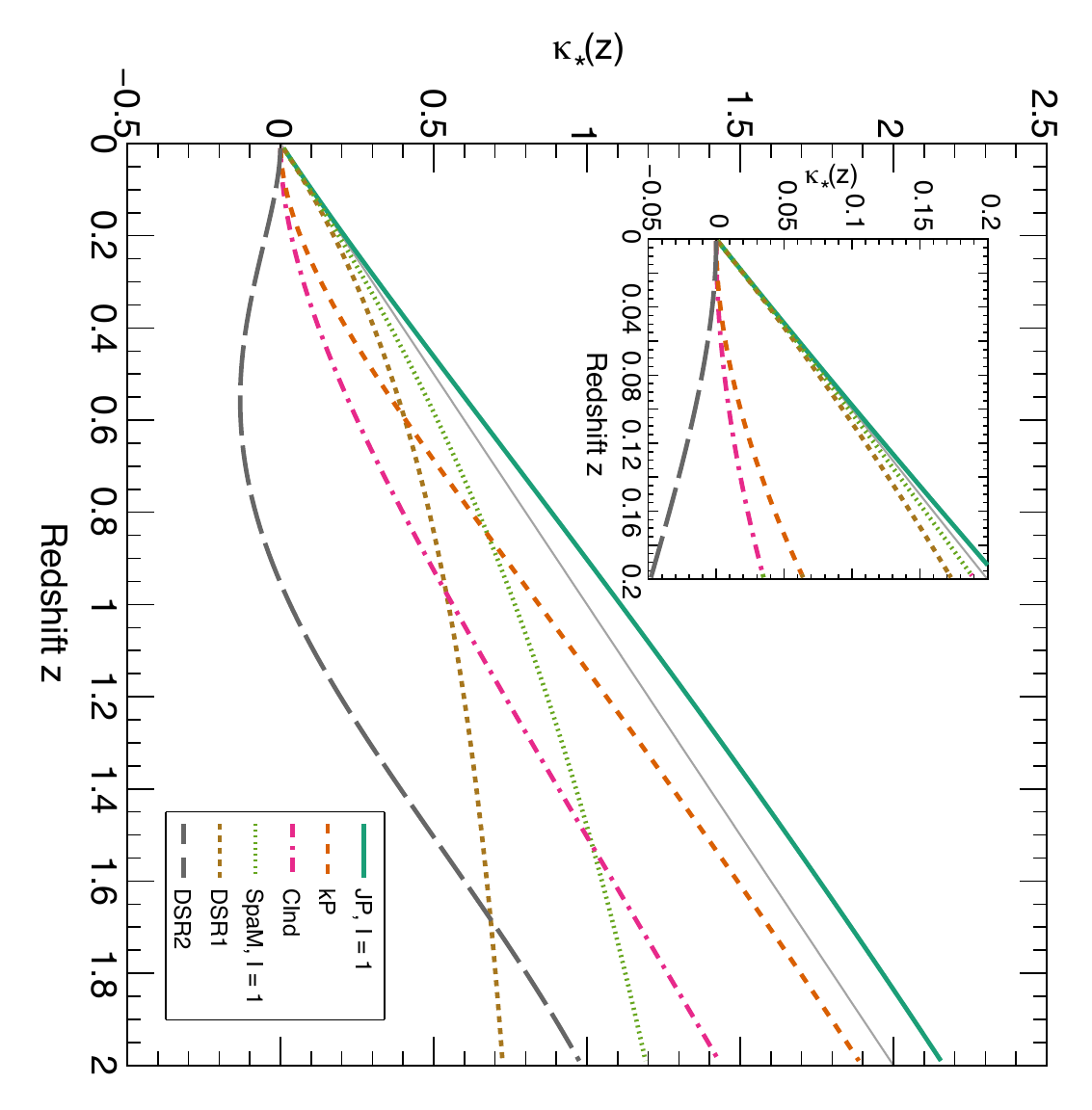}
	\caption{The dimensionless lag-redshift relation $\kappa_*$  as a function of redshift for values up to $z=2$, for the six linear ($I = 1$) models considered in the analysis (Sec.~\protect\ref{sec:results}). The inset plot shows a detailed view of the models at low redshifts. The thin gray solid line corresponds to the identity $\kappa_*(z) = z$ and is shown only as a reference.}\label{fig:Kappa}
\end{figure}

Fig.~\ref{fig:Kappa} shows the function $\kappa_*(z)$, for the different models. It can be observed that high redshift trends tend to be similar between all models except for model \eqref{eq:RekaP} and \eqref{eq:SpaM} which exhibit a flatter trend. At low redshifts ($z < 0.2$), \eqref{eq:JP}, \eqref{eq:SpaM}, and \eqref{eq:RekaP} give very similar lag-redshift relations, while the other models are significantly lower. Model \eqref{eq:DSR2} is the only one that gives negative values for $z < 0.95$, with a decreasing trend up to $z \approx 0.55$. 

In the following, we will compare the sensitivity of data analysis for the study of time delays induced by the different models listed above. We will also check how important it is to choose the correct time-delay model when implementing the search for QG-induced time lags. 

\section{Simulations and analysis} \label{sec:dataanalysis1}
To assess the impact of different modified dispersion models on time-delay measurement, we focus here on variable or transient extragalactic sources, located at different redshifts. The methodology in use is similar to the one described in detail in \cite{Bolmont_2022}, and we emphasize only the most important details in the next two sub-sections. For each source, flaring AGN or GRB, a simulated data set mimicking actual data is generated (see Sec.~\ref{sec:Simulations}), including, or not, an energy dependent time delay, obtained from the various MDRs highlighted in the previous section. Then, a likelihood analysis technique is used to reconstruct the delay~(see Sec.~\ref{sec:Analysis method}).

Pulsars are not considered in the present study, as all of those detected are located at $z = 0$. Therefore, they could not be used to discriminate between the models in use here, since those models converge to the same predicted time lags at the $z \to 0$ limit. However, this also means pulsars are independent of any particular lag-redshift model. In addition, their short pulsation periods offer very strict constraints on the emission times. Furthermore, unlike AGNs and GRBs, whose flares and bursts are random, pulsar behavior is predictable. Finally, they can be observed with very high statistics on a wide energy range. For all these reasons, they are undoubtedly valuable candidates for LIV searches and must be considered in future population studies.

\subsection{Simulations}
\label{sec:Simulations}

The simulated datasets are meant to reproduce two flaring AGNs detected by MAGIC and H.E.S.S., respectively Mkn~501 \cite{Albert:2007zd, Mart_nez_2009, MAGIC:2007etg} and PKS~2155-304 \cite{Aharonian:2007ig, Abramowski2011}. Two GRBs are included as well: GRB~190114C detected by MAGIC \cite{MAGIC:2019lau, MAGIC:2019irs, Acciari_2020}, and GRB~090510 detected by \textit{Fermi}-LAT \cite{2009Natur.462..331A, Vasileiou_2013}. This combination allows an almost uniform distribution of sources over a range in redshift spanning from $z = 0.034$ (Mkn~501) to $z = 0.903$ (GRB~090510). Each simulated dataset consists of $1000$~realizations of the same burst or flare. For each source, two datasets are produced: one with the same number of simulated events as in the actual observation, and a second one with a factor $10$ enhancement of the event statistics. The latter simulations are referred to as ``boosted simulations'' in the following. The improvement of photon statistics by one order of magnitude is expected for future generation facilities, such as CTAO~\cite{CTAO_performance}, which are designed to allow an order of magnitude improvement of the sensitivity as compared to current instruments. In addition, simulations allow to inject a propagation lag following any of the models highlighted in Sec.~\ref{sec:theory}: \eqref{eq:JP}, \eqref{eq:RekaP}, \eqref{eq:curvInd}, \eqref{eq:SpaM}, \eqref{eq:DSR1} and \eqref{eq:DSR2}.

\subsection{Analysis method}
\label{sec:Analysis method}

The analysis method closely follows the one described in \cite{Bolmont_2022}. Namely, a negative log-likelihood function is minimized in order to find the optimal lag parameter $\lambda$ to fit the distribution of gamma-like events over time and energy of each simulated dataset. The used probability density function depends on energy $E$ of the photon in the rest frame of the detector and time $t$. It is defined as: 

\begin{equation}\label{eq:realistic_pdf}
\begin{aligned}
 \frac{d P}{d E d t}  = w_s\ \frac{ F_s\left(E, t ; \lambda\right)}{N_s^{\prime}} 
 +\sum_k w_{b, k}\ \frac{ F_{b, k}\left(E\right)}{N_{b, k}^{\prime}}\,,
\end{aligned}
\end{equation}
where $w_s$ is the proportion of signal events and $w_b = \sum_k w_{b,k}$ is the proportion of background events, taking into account baseline photons and hadrons. $F_s$ and $F_{b,k}$ are respectively the distribution of signal and background events which depend on 
\begin{equation}
\begin{aligned}
|\lambda| = \frac{\Delta t}{E_{01}-E_{02}} \times \frac{1}{\kappa_*(z)} = \frac{1}{H_0 \Lambda}.
\label{eq:lambdatoLambda}
\end{aligned}
\end{equation}
 Positive $\lambda$ is interpreted as subluminal while negative $\lambda$ is interpreted as superluminal. The different models described in Sec.~\ref{sec:theory} only modify the $\kappa_*(z)$ function, which allows those models to be transparently taken into account in the computation of the best estimate for $\lambda$. Instrumental response functions and nuisance parameters are neglected in both simulations and analysis, in order to focus on the difference between the theoretical models of $\kappa_*(z)$. The impact of the systematic uncertainty for the current generation of Cherenkov telescopes has been studied in \cite{Bolmont_2022}. 

This uncertainty typically weaken the constraints by a factor of the order of $2$ independently of the theoretical model. 
For \textit{Fermi}-LAT the systematic uncertainties, as provided in \cite{Vasileiou_2013}, would reduce the limits by 10\,\%. The log-likelihood $L_S$ for each source is then constructed as the sum of the logarithm of the probability density function of each individual event: 
\begin{equation}
\begin{aligned}
L_S(\lambda) = - 2\sum_{i} \log \left( \frac{dP}{dE dt}(E_{i},t_i);\lambda \right)\,,
\end{aligned}
\end{equation}
and is summed over sources to obtain the total likelihood: 
\begin{equation}
\begin{aligned}
L_{\mathrm{comb}}(\lambda) = \sum_{\mathrm{all\ sources}} L_S(\lambda).
\end{aligned}
\end{equation}
In the case where the fitted $\lambda$ (corresponding to the minimum of the $-2\log$ curve and therefore called $\lambda_{\mathrm{min}}$) is compatible with zero and data are Gaussian distributed, a $95\,\%$ confidence level limit called $\lambda_{\mathrm{lim}}$ can be extracted by solving the equation~\cite[Table 40.2]{PhysRevD.110.030001}: 
\begin{equation}
\begin{aligned}
L_{\mathrm{comb}}(\lambda_{\mathrm{lim}}) = L_{\mathrm{comb}}(\lambda_{\mathrm{min}}) + 3.84.
\end{aligned}
\end{equation}
The obtained $\lambda_{\mathrm{lim}}$ can then be converted to the corresponding limit on the quantum gravity scale $\Lambda$ using Eq.~\eqref{eq:lambdatoLambda}.

\begin{figure*}[!htb]
	\centering
	  \includegraphics[width=0.49\textwidth]{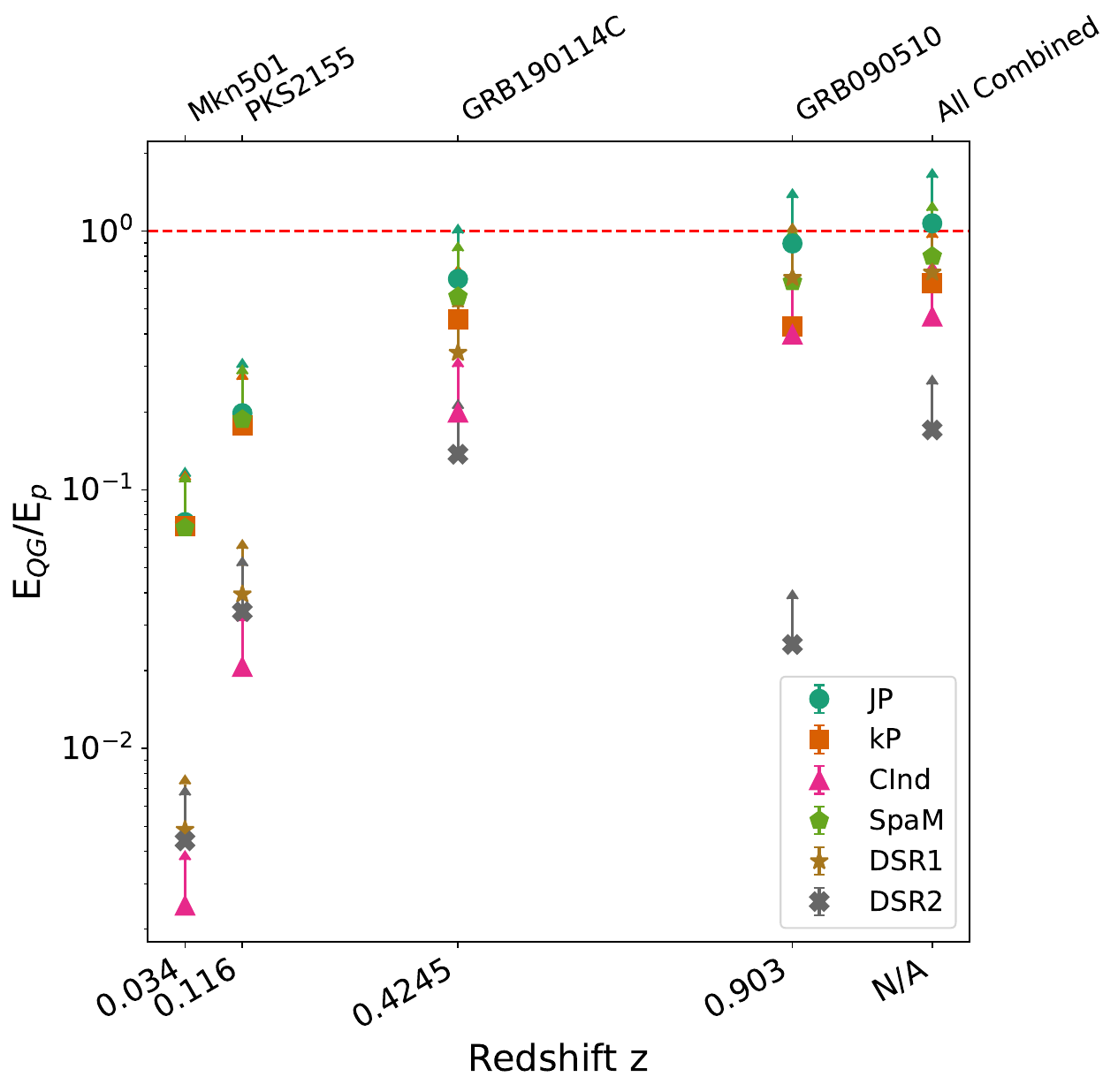}
   \includegraphics[width=0.49\textwidth]{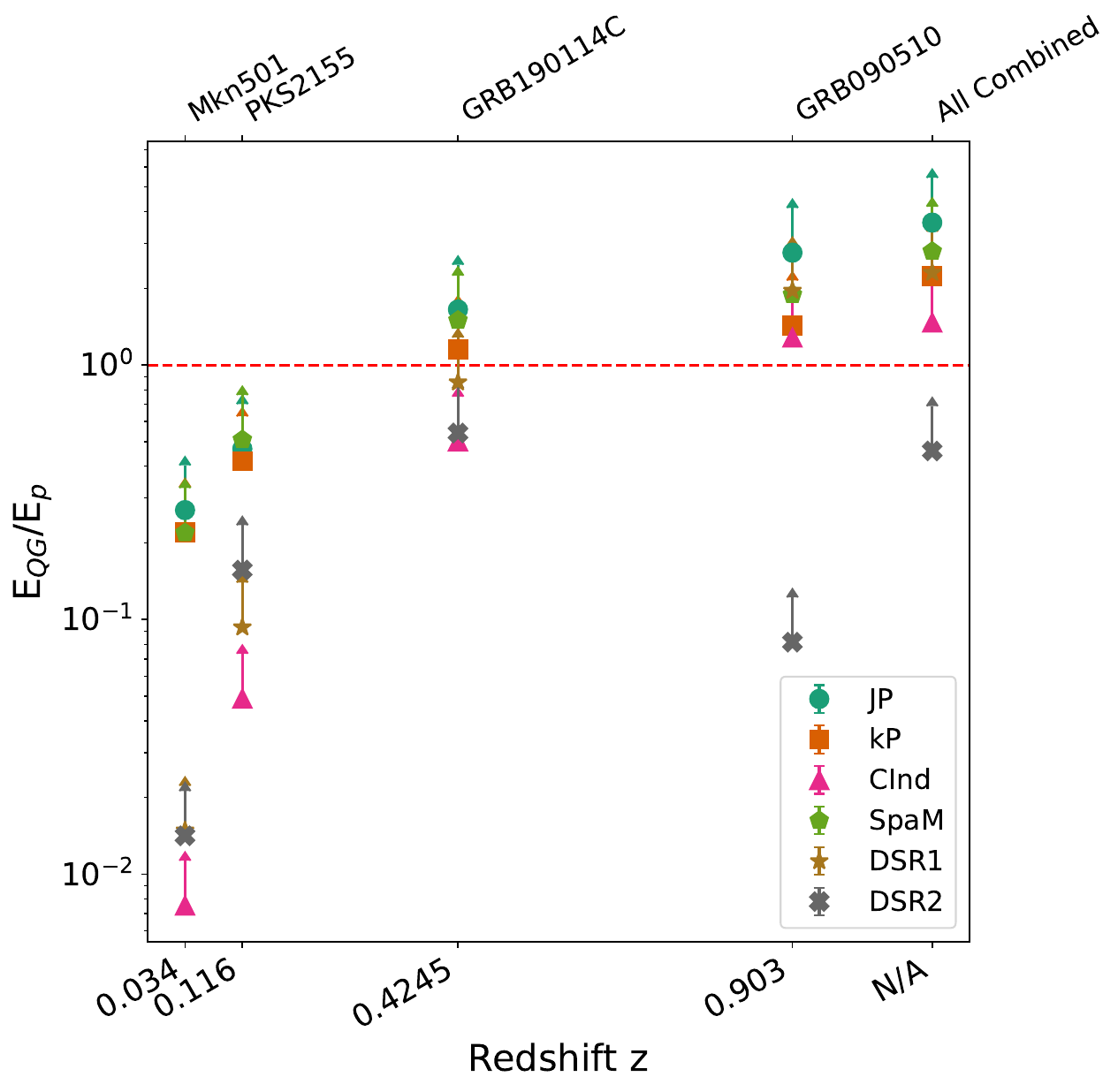}
	\caption{Lower limit on energy scale $\Lambda$ as a function of redshift when no lag is injected. The last point on the right shows the combination of all the simulated sources. The left panel shows results with standard simulations while the right panel shows results with boosted simulations, i.e. simulations where the event statistics have been increased by one order of magnitude. As stated in the text, these limits are meant to be compared to each other, emphasizing the differences introduced by various lag-redshift models. They were obtained neglecting instrument response functions and systematic uncertainties.}
 	\label{fig:Simulations_lim}
\end{figure*}

\section{Results}\label{sec:results}

The results are obtained in two different cases. First, datasets are simulated with no lag injected (Sec.~\ref{subsec:lagfree}). This is equivalent to assuming there is no QG-induced time delay. Second, a lag is injected (Sec.~\ref{subsec:withliv}), which would correspond to the case where $\Lambda = E_P$ in the \eqref{eq:JP} model.

\subsection{Lag-free simulations}\label{subsec:lagfree}

The quantum gravity scale $\Lambda$ is constrained using simulations with no injected lag, mimicking observations where a delay would not be statistically significant. In that case, lower limits are derived following the procedure described in the previous section (\ref{sec:Analysis method}). In order to assess how the constraint changes with the redshift for the different benchmark models listed in Sec.~\ref{ssec:MDRsspecific}, we perform the analysis on individual sources listed in \ref{sec:Simulations}. In addition, we also compute for each model the limit obtained by combining all sources in the sample.

Fig.~\ref{fig:Simulations_lim} shows the obtained limits for the various sources and their combination. The limits suggest the lag-redshift models can be separated into several classes. 
Models \eqref{eq:DSR1}, \eqref{eq:DSR2}, and \eqref{eq:curvInd} give significantly lower limits (i.e. less constraining) for nearby sources (typically $z<0.4$) than \eqref{eq:JP},  \eqref{eq:SpaM}, and \eqref{eq:RekaP}. So far, most of the bright and variable AGNs have been observed at these redshifts by the current generation of IACTs. Therefore, low-redshift AGNs would be disadvantaged for constraining or detecting QG-induced time delays for models \eqref{eq:DSR1}, \eqref{eq:DSR2} and \eqref{eq:curvInd}. 

At the same time, model \eqref{eq:DSR2} reduces constraining ability both at redshifts $z\sim0$ and $z\sim1$. In that sense, GRB~190114C is at nearly optimal redshift ($z=0.4245$) to constrain model \eqref{eq:DSR2}.
It is interesting to note that for this particular model, the constraint of GRB~090510 is poor compared to the ones based on the sources with smaller redshifts. The systematic uncertainty resulting from the choice of the time-delay model can be reduced by combining sources at various redshifts.
Although different models can lead to one order of magnitude of difference in the lower limit, a combination of appropriately sampled sources in terms of redshift enables reducing the difference between models. 
For the boosted dataset, the situation remains the same with higher lower limits overall. It is interesting to note that an upper limit can exceed the Planck energy for a particular model, while being lower than it with another model. 
Another interesting finding is that \eqref{eq:JP} gives systematically higher limits than the other models. This is explained by the fact the time lag (Eq.~\ref{eq:DtKappa}) for \eqref{eq:JP} is larger than the lags for the other models (as seen on Fig.~\ref{fig:Kappa}).

Focusing on GRB~090510, a significant discrepancy between the results from our simulations and the ones obtained from actual data published in \cite{Vasileiou_2013} can be noticed. The one thousand simulations give a scale lower than the Planck scale in all models. This discrepancy is too large to be explained by a simple statistical effect. Moreover, our simulation results are compatible with the uncertainties given in \cite{Vasileiou_2013} for the Pair View (PV) and Sharpness-Maximization Method (SMM), while reducing the important bias observed for GRB~090510 in the same paper. Also, we notice that the likelihood method in that paper gives results with an uncertainty reduced by a factor of ten compared to the PV and SMM methods for GRB~090510, while this is not the case for the other GRBs of the paper. It seems to us that only a new analysis of the same dataset with the same software we use would be able to solve this discrepancy. Such an investigation is beyond the scope of the present paper.

\subsection{Simulations with an injected LIV time delay}\label{subsec:withliv}

\begin{figure}[!t]
	\centering
   \includegraphics[width=3.4in]{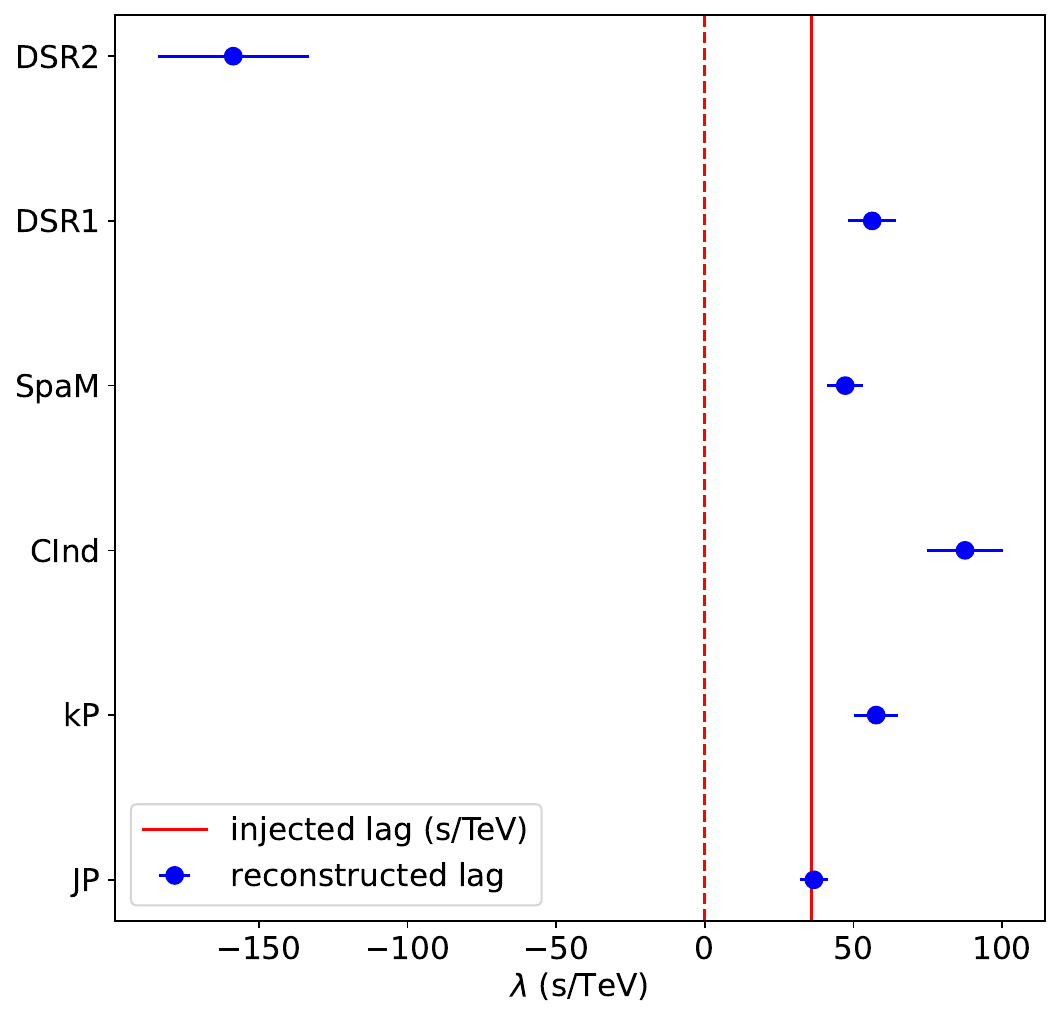}
	\caption{Reconstructed lag from the combination of all the sources (blue points) for an injected lag corresponding to $\Lambda\simeq E_P$ for the \eqref{eq:JP} model (red solid line). The null hypothesis case ($\lambda$ = 0\,s\,TeV$^{-1}$) is shown with a red dashed line.}
	\label{fig:lag_injected}
\end{figure}

In order to assess the impact of the usage of a wrong model for the detection of a QG time delay effect, we used boosted simulations with an injected lag $\lambda = 36$~s\,TeV$^{-1}$ corresponding to $\Lambda\simeq E_P$ for the \eqref{eq:JP} model. Then, using our analysis software, we reconstructed the lag $\lambda$ by using the various lag-redshift models on the combination of all the sources. The results are shown in Fig.~\ref{fig:lag_injected}. While, as expected, the reconstructed lag corresponds well to the injected lag for the \eqref{eq:JP} model, the use of a wrong model significantly biases the reconstruction of the lag, leading to an overestimation of this parameter for all models except \eqref{eq:DSR2}, for which it is underestimated. The latter case is particularly interesting, since the sign of the reconstructed lag is inverted, possibly leading to a misinterpretation between superluminal or subluminal models. In addition, it is important to note that even if the lag is shifted compared to its true value, none of the models obtain a lag compatible with zero. In other words, a real lag will always be detected, even when using a wrong model. However, using the wrong model would inevitably lead to an incorrect value of $\Lambda$. In this context, it is interesting to check whether the data themselves can help discriminating between the different models. 

\begin{figure}[!t]
	\centering
\includegraphics[width=3.4in]{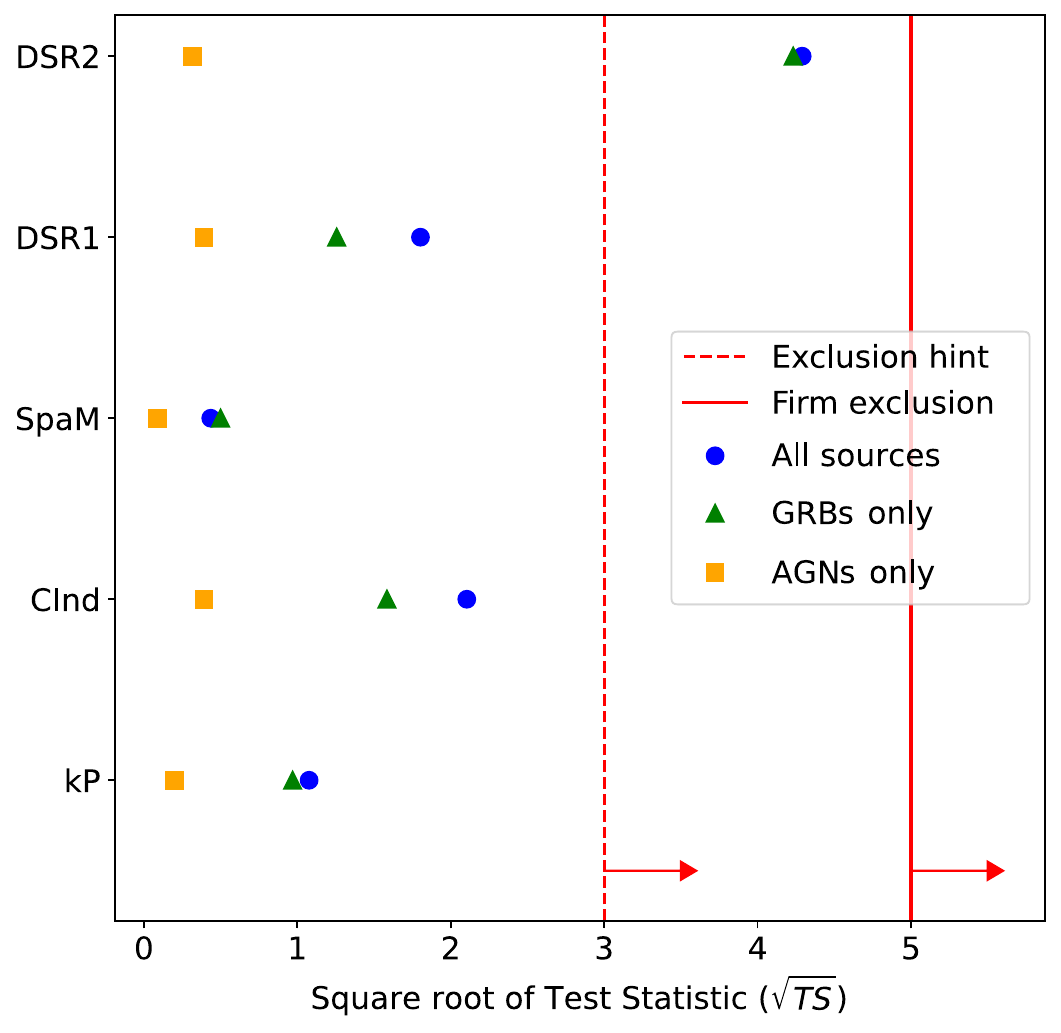}
	\caption{Statistical tests for the various lag-redshift relation models, obtained with AGN only, GRB only, and a combination of the full sample. The results are obtained from the boosted simulation, injecting a lag $\lambda = 36$~s\,TeV$^{-1}$. Red dashed and solid lines correspond respectively to three and five standard deviations, representing an exclusion hint and a firm exclusion of a given model.}
	\label{fig:Model_discrimination}
\end{figure}

To test this possibility, we produced a sample with the \eqref{eq:JP} model with a lag of $\lambda = 36$\,s\,TeV$^{-1}$. Then, we calculated the likelihood using the \eqref{eq:JP} model and compared it with the likelihoods obtained with the other models. 
We performed a likelihood ratio statistical test, using the boosted simulations, by computing the square of the $L_\mathrm{comb}(\lambda_\mathrm{min})$ ratio obtained from the \eqref{eq:JP} model and each tested model. The statistical test was performed using only AGNs (Mkn~501 and PKS~2155-304), only GRBs (GRB~190114C and GRB~090510), and combining all the sources. The results are shown in Fig.~\ref{fig:Model_discrimination}. No model can be firmly excluded at a $5\sigma$ level with the sample used, but the \eqref{eq:DSR2} case reaches $4.3\sigma$, giving a strong hint of exclusion. While the combination is clearly driven by the GRB sample for models \eqref{eq:DSR2}, \eqref{eq:RekaP} and~\eqref{eq:SpaM}, the picture is less clear for models \eqref{eq:DSR1} and \eqref{eq:curvInd}. The presence of AGNs in the sample clearly helps in discriminating those two models. This can be explained by the fact these models mostly differ at low redshifts. It can be seen that \eqref{eq:RekaP} and \eqref{eq:SpaM} are poorly excluded due to the similarity of their trend compared to the \eqref{eq:JP} model. It is particularly interesting to note that for~\eqref{eq:SpaM} the AGNs of the sample do not contribute to improving the combined limit, given its similarity to~\eqref{eq:JP} at low redshift. Our analysis shows that, depending on the tested model, discrimination can be obtained easier at low or high redshift. Thus, the major conclusion is that a good sampling in terms of redshift is very important to discriminate between different models.

\section{Discussion and outlook} \label{sec:DiscOut}

The search for quantum gravity effects is one of the most fascinating endeavors in nowadays physics research. At all scales, the sensitivity of measurements approaches a level which allows us to realistically probe Planck scale physics. Hence, insights to the question if gravity is quantized or not, and if it is, how it affects the propagation of photons on spacetime, become attainable.

There are numerous models in the literature which effectively describe the propagation of photons on quantum spacetime and predict a time delay on the basis of MDRs, the oldest and the most prominent one being the \eqref{eq:JP} model \cite{Jacob:2008bw}. So far, most experimental studies of MDRs focused on this model used it to derive constraints on the quantum gravity / new physics beyond general relativity and the standard model scale $\Lambda$.

In the present article, a new, very general parameterization of MDRs is presented. It characterizes the redshift dependence of the first non-trivial deviations from general relativity by a set of parameter functions $B_m(t)$, introduced in Eq.~\eqref{eq:MDRpertGen2}. This representation of MDRs directly translates into a parameterized form of the time delay Eq.~\eqref{eq:DT2} and of the light distance function Eq.~\eqref{eq:kappaB}. The most prominent models already presented in the literature (\eqref{eq:JP}, \eqref{eq:curvInd}, \eqref{eq:DSR1}, \eqref{eq:DSR2}), and new ones such as \eqref{eq:SpaM} and \eqref{eq:RekaP}, can be obtained by choosing the values of $B_m$ accordingly.

From the MDRs obtained in the models, simulated datasets were generated based on observations of 
four events: GRB~090510 (\textit{Fermi}-LAT), GRB~190114C (MAGIC), and PKS~2155-304 and Mkn~501 (H.E.S.S.)
in two situations: with no injected lag (no quantum spacetime effect) and with an injected lag corresponding to an effect at the Planck scale in the \eqref{eq:JP} model. The lag was then reconstructed with a log-likelihood technique for each model and each source, as well as for each model when combining the results for all sources.

We found that the different models in the literature, and the new model \eqref{eq:SpaM} which we added for the analysis, lead to different bounds on the QG energy scale (Fig.~\ref{fig:Simulations_lim}) when no lag is injected. As an example, for the most sensitive combined analysis of all events, we found that the \eqref{eq:DSR2} model is compatible with a bound of $E_{QG} > 0.16\ E_{P}$, while the \eqref{eq:JP} model already leads to a bound of the order of $E_{P}$. 
That means that, due to different lag-redshift dependencies between these models, we are capable of exploring an order of magnitude larger parameter space of $E_{QG}$ for \eqref{eq:JP} than for the \eqref{eq:DSR2} model.

When a lag was injected into the data, assuming a QG effect at the Planck scale, and then reconstructed using the different models, we clearly found that, depending on the model which underlies the analysis, different values of the measured time lag are obtained from the same dataset. However, all models are consistent with each other, and incompatible with the no time lag hypothesis. Thus, if a time lag is present in real observations, the genuine feature could be detected independently of the model. However, the magnitude of the lag is clearly model dependent. It has to be stressed here that the detection of a significant time lag in actual observations would require a proper interpretation. In particular, time delays can be introduced by internal source mechanisms and they would need to be separated from quantum gravity induced propagation effects. 

So far, no time delay at GeV or TeV energies has been observed. In case of a clear detection of a time lag, our new parameterization opens the way to fit the coefficients $B_m(t)$ to an observation and to compare the fitted coefficients to the ones derived for different models. The development of such a reconstruction of the parameters from observations is in progress. 

Since we find that different models can lead to different conclusions about the bounds on the new physics or quantum gravity energy scale $\Lambda$, we encourage future experimental searches to use models beyond the original~\eqref{eq:JP} model, such as the ones we analyzed here. As an example, the limits obtained by the LHAASO collaboration with data from GRB~221009A \cite{LHAASO:2024lub} would be lower by a factor of 8 if the model \eqref{eq:curvInd} would be considered instead of the \eqref{eq:JP} model (see Fig. \ref{fig:Kappa}).

The data analysis code used in~\cite{Bolmont_2022} has been updated in order to be able to use any kind of model. For the future, any new model of interest can easily be tested against observations. Ideally, the general parameterized model Eq.~\eqref{eq:DT2} could be used as well. 
In the long run, this will hopefully lead to the identification of the most viable MDR model.
    
\begin{acknowledgments}
    
    The authors would like to acknowledge networking support by COST Actions CA18108 QGMM (Quantum Gravity in the Multi-Messenger era, \url{https://qg-mm.unizar.es/}) and CA23130 BridgeQG (Bridging high and low energies in search of Quantum Gravity, 
    \url{https://www.cost.eu/actions/CA23130/}.
    They also would like to thank the Institute for Fundamental Physics of the Universe (IFPU, \url{https://www.ifpu.it/}) for hosting the workshop ``Astrophysical searches for quantum-gravity-induced time delays'' where ideas important to this work were developed.
    
    Moreover, the authors would like to gratefully thank the colleagues from the University of Zaragoza QuGraPheno research group, in particular J.M.~Carmona, J.L.~Cort\'es and M.A.~Reyes, for their important and insightful comments.
    
    C.P. acknowledges the financial support by the excellence cluster QuantumFrontiers of the German Research Foundation (Deutsche Forschungsgemeinschaft, DFG) under Germany's Excellence Strategy -- EXC-2123 QuantumFrontiers -- 390837967 and was funded by the Deutsche Forschungsgemeinschaft (DFG, German Research Foundation) - Project Number 420243324.

    T.T., J.S., and C.P. acknowledge the support from the University of Rijeka through project uniri-iskusni-prirod-23-24. T.T. and J.S. acknowledge the support from the Croatian Science Foundation (HrZZ) Project IP-2022-10-4595.

    C.Pl. and S.C. acknowledge funding from the French \textit{Programme d’investissements d’avenir} through the Enigmass Labex.

    S.W. Acknowledges the support from the following agencies and institutions: Natural Sciences and Engineering Research Council of Canada, Calcul Québec, Centre de Recherche en Astrophysique du Québec: un regroupement stratégique du Fonds de Recherche du Québec Nature et Technologies (FRQNT).

\end{acknowledgments}



\vspace{1em}

The work presented in the present article was first initiated by C.P., T.T., S.C. and J.B. during discussions at various QGMM COST Action events. C.P. formulated the theory, derived and extracted the time delay formulas from the modified dispersion relations. J.B. and S.C. implemented them in the software and S.C. produced the results presented in the paper, as well as the plots. S.C., C.P., J.B. and T.T. were equally involved in the editing of the manuscript.

The software is being developed with contributions from all the authors, who (except for C.P.) are members of either of the H.E.S.S., MAGIC, VERITAS or LST collaborations.


\bibliography{TLGRB}

\end{document}